\documentclass[pra,letterpaper,showpacs,superscriptaddress]{revtex4}
\usepackage{graphicx,amsmath,amssymb,amsfonts,color,pstricks-add}

\begin{document}

\title{On the possibility of Casimir repulsion using Metamaterials}

\author{F S S Rosa}
\affiliation{Theoretical Division, Los Alamos National Laboratory, Los Alamos, NM 87545, USA}

\begin{abstract}
It is well known that the Casimir force between two half-spaces is dictated by their electromagnetic properties.  In particular, when one of the half-spaces is mainly metallic or dielectric and the other is mainly magnetic, it is possible to show that the force is repulsive. This has attracted lots of interest towards the study of metamterials (MMs) in the context of Casimir effect, as their magnetic activity might help bring the idea of Casimir repulsion from the theoretical realm to experimental verification. In this paper we investigate the possibility of repulsion when the MM magnetic permeability is given not by a Drude-Lorentz behavior, but by a model put forward by Pendry et al. \cite{Pendry}. After introducing the model and deriving the necessary formulas, we show that it is impossible to achieve repulsion with such a model and present a qualitative discussion of why this is so. 
\end{abstract}

\maketitle

\section{Introduction}

The last decade has witnessed an increased interest in Casimir physics \cite{Casimir,reviewsCasimir} thanks to improved precision measurements \cite{casimirexperiments} of the force between material objects separated by micron and sub-micron gaps. In a few words, the Casimir force may be thought of as a consequence of changing the vacuum fluctuations spectrum through the insertion of material boundaries. While the Casimir force offers new possibilities for nanotechnology, such as actuation mediated by the quantum vacuum, it also presents some challenges, as micro and nanoelectromechanical systems (MEMS and NEMS) may stick together and cease to work due to the attractive nature of van der Waals and Casimir forces. Recent years have also witnessed a huge activity in the development of metamaterials (MMs) \cite{reviewsMM}, boosted by the possibility that such engineered media may give rise to novel optical properties at selected frequency ranges, including negative refraction \cite{negative_refraction}, perfect lensing \cite{perfect_lens}, and cloaking \cite{cloaking}, among others. Generally speaking, metamaterials are made of micro and nanostructures carefully designed to collectively present a particular electromagnetic feature. They are able to produce the aforementioned striking phenomena, inaccessible with natural materials, due to the significant magnetic activity built into them, starting at microwave frequencies and going all the way up to the optical range. 

Substantial magnetic activity at high frequencies is exactly what brings Casimir physics and metamaterials close together. It is known for a long time that the attractive character of Casimir forces is not universal, that it actually depends crucially on the geometry and composition of a given setup. This naturally leads to speculations about the feasibility of customized systems displaying repusive Casimir forces, or maybe strongly reduced attraction, thus providing a so desired anti-``stiction" effect. An interesting possibility of flipping attraction into repulsion was first demonstrated by Boyer \cite{Boyer}, where he showed that a perfect conducting plate repels a perfect magnetic one with vacuum in between. The requirement of having a perfect magnet was relaxed in subsequent works \cite{Klich}, where it was shown that non-ideal materials could give rise to repulsion (provided one was mainly electric and the other mainly magnetic), but even this more realistic scenario is constrained by the absence of natural strong magnets at high frequencies\footnote{There are plenty of natural strong magnets at low frequencies, but they become magnetically inert at frequencies well below $c/d$, where $d$ is the distance between the half-spaces.}. However, recent developments in nanofabrication have resulted in metamaterials with magnetic response in the visible range of the electromagnetic spectrum \cite{MMs}, fueling the hope for Casimir repulsion \cite{repulsion}. The expectation is that, by tuning this magnetic response to the right frequency range and making it strong enough, one could produce an experimentally measurable Casimir repulsion between, say, a MM slab and a thin metallic plate, or at least a significantly reduced attraction. 

The major issue about Casimir repulsion and, more generally, about any degree of Casimir force control is that the Casimir interaction between real dispersive materials is a broadband frequency phenomenon, as shown by the Lifshitz formula expressing the force between two semispaces as an integral over all (imaginary) frequencies with an exponential cut-off $c/d$ \cite{Lifshitz}. In previous works \cite{usPRL,usPRA} we have shown that this is a formidable obstacle on the route to Casimir repulsion, as typically the magnetic response of a MM is restricted to a narrow frequency band. In this work we would like to discuss an additional issue, briefly touched in \cite{usPRL} (see also \cite{Intravaia}), concerning the modeling of magnetic activity in a given metamaterial. As we shall see below, this is a very important discussion since the Casimir force depends crucially on the choice of the proper model.

\section{Metamaterials: a simple model}

As our first step, we want to present a simple yet effective procedure of how to characterize the magnetic activity of a metamaterial. This is better illustrated by tackling a specific example, so let us consider the metamaterial depicted in Fig. 1 \cite{Pendry}. It consists on a periodic array of double layered conducting cylindrical sheets, each one being characterized by its radius $r$, separation $s$ between the sheets and length $L$. The layers themselves have some structure - they are tinily split apart on opposite sides, in such a way to prevent currents of completing an entire loop (see fig. 2). The array itself has period $b$, in addition we assume that the cylinders are long ($r/L \ll 1$) and that there is little space between the two layers ($s/r \ll 1$) of a given cylinder. 
\begin{figure}[h]
\begin{minipage}{14pc}
\scalebox{0.35}{\includegraphics{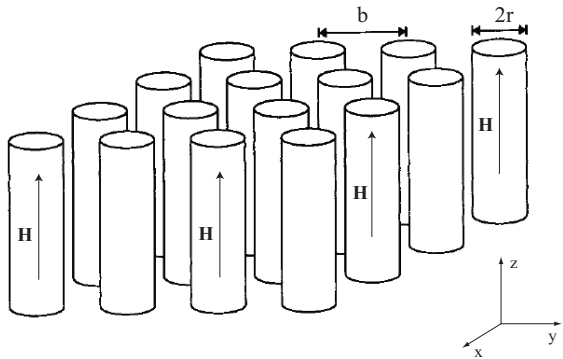}}
\caption{\small Sketch of the array of cylinders. The distance from center to center (the period) is $b$ and each cylinder has radius r. For the sake of clarity, we don't show the structure of each cylinder in this figure.}
\end{minipage}\hspace{100pt}
\begin{minipage}{14pc}
\psarc[linewidth=.3mm]{-}(2,0){1.4}{10}{350}
\psarc[linewidth=.3mm]{-}(2,0){1.7}{190}{530}
\psarc[linewidth=.2mm]{<-}(2,0){1.25}{60}{120}
\psarc[linewidth=.2mm]{->}(2,0){1.85}{60}{120}
\psline[linewidth=.2mm]{->}(2,0)(3,-1)
\psline[linewidth=.2mm]{->}(1.5,-0.5)(1,-1)
\psline[linewidth=.2mm]{->}(0.3,-1.7)(0.8,-1.2)
\pscircle[linewidth=.2mm](1.5,0.3){0.16}
\pscircle[linewidth=.2mm](2.5,0.3){0.16}
\psdot(1.5,0.3)
\psdot(2.5,0.3)
\rput(1.1,0.3){\small \bf H}
\rput(2.9,0.3){\small \bf H}
\rput(2,0.9){J}
\rput(2,2.1){J}
\rput(2.7,-0.45){r}
\rput(0.9,-1.1){s}
\vspace{60pt}
\caption{\small Structure of each cylinder, seen from the top.}
\end{minipage}
\end{figure}
By turning on an external magnetic field ${\bf H}(t) = H(t) \hat{z} = H_0 \mathrm{e}^{-i\omega t} \hat{z}$ parallel to the cylinders we induce circulating currents on them, that therefore induce a new magnetic field ${\bf H}_1$\footnote{In principle, this new field induces another current, which in turn induces another field and so on and so forth. These subsequent fields and currents, however, are very small and therefore neglected in this approximate model.}. The total field is therefore
\begin{equation}
\label{TotalField}
{\bf H}_T({\bf x}) = {\bf H} + {\bf H}_I({\bf x}) = (H + H_I({\bf x}))\hat{z} \, ,
\end{equation}
where ${\bf x}$ is the position where the field is being evaluated and $H_I({\bf x})$ is still to be determined. In this approximate model, our first assumption is that $H_1$, despite having different expressions inside and outside a given cylinder, is space independent in each of these two regions. Inside of a given cylinder, the field may be naturally divided into two contributions: the one that originates from that cylinder itself and another that comes from all the other cylinders combined, recalling here that the cylinders are finite in length and as such they ``leak'' some magnetic field to their surroundings. The first contribution is readily expressed in terms of the induced current per unit of length $J$ with the help of Amp\`ere's law, resulting in $H_I^{(1)} = J/c$. This of course holds exactly only for infinite cylinders, but is an excellent approximation for cylinders sufficiently long. For the second contribution we shall present an heuristic argument, based on the conservation of the number of field lines. We know that the field strength in a given location is roughly proportional to the density of field lines in that location, so the first contribution gives 
\begin{equation}
\label{FieldLines}
H_I^{(1)} = \kappa \rho_{\rm in} = \frac{\kappa N}{{\pi} r^2} \, ,
\end{equation}
where $N$ is the number of lines, $\rho_{\rm in}$ is the density of lines and $\kappa$ is a proportionality constant. By following the same reasoning, in order to get  the second contribution we must find the density of field lines ``leaked" by all the cylinders, which is simply the density $\rho_{\rm out}$ ``leaked" by one cylinder times the number of cylinders $N_{\rm cyl}$
\begin{equation}
\label{Heuristic}
|H_I^{(2)}| = \kappa \rho_{\rm out} N_{\rm cyl} = \kappa \frac{N}{A} \frac{A}{b^2} = \kappa \frac{N}{b^2} \, ,
\end{equation}     
where $A$ is the (infinite) area of a plane perpendicular to the cylinders and we used the fact that a given (finite) cylinder leaks all of its lines to its surroundings. From (\ref{Heuristic}) we get 
\begin{equation}
\label{Ratio}
\frac{|H_I^{(2)}|}{H_I^{(1)}} = \frac{\pi r^2}{b^2}\, ,
\end{equation}    
and, knowing that $H_I^{(2)}$ and $H_I^{(1)}$ have opposite signs, we may use our previous results to rewrite the total field inside a cylinder as
\begin{eqnarray}
\label{CylinderArrayIn}
H_T^{\rm in} = H + \frac{1}{c} J - \frac{1}{c}\frac{\pi r^2}{b^2}J \, .
\end{eqnarray}
Since the field outside the cylinders may be worked out in a similar way, we shall just quote the result 
\begin{equation}
\label{CylinderArrayOut}
H_T^{\rm out} = H - \frac{1}{c}\frac{\pi r^2}{b^2}J  \, ,
\end{equation}
and point out that the only difference is that $H_I^{(1)}$ vanishes in this case. The key advantage of expressions (\ref{CylinderArrayIn}) and (\ref{CylinderArrayOut}) is that they are written in terms of the current density $J$, to what now we turn our attention. Each cylinder may be viewed as an RC circuit, and application of Faraday's law in one of them gives 
\begin{eqnarray}
\label{Faraday}
{\cal E} = 2\pi r \alpha J -\frac{J}{i\omega C} = - \frac{\pi r^2}{c}\frac{\partial}{\partial t} \left[ H_0 + \frac{1}{c} J - \frac{1}{c}\frac{\pi r^2}{b^2}J \right] \; ,
\end{eqnarray}
where where $\alpha$ is the resistivity of the cylinder sheets and ${\cal C}$ is the effective circuit capacitance also per unit of length. By assuming that it evolves harmonically in time, i.e., that $J=J_0 e^{-i\omega t}$, we get
\begin{equation}
\label{Current}
J = \frac{ i \omega H}{\frac{2\pi\alpha c}{\pi r} - \frac{c}{i \omega {\cal C} \pi r^2} - \frac{i\omega}{c} \left(1- \frac{\pi r^2}{b^2}\right)} \, ,
\end{equation}
which might be used in (\ref{CylinderArrayIn}) and (\ref{CylinderArrayOut}) to give explicit expressions to the fields. We are not, however, so interested in these absolute expressions as we are in their spatial average, since the latter presumably gives a better account of effective properties, specially at large scales. Starting from the integral form of Faraday's and Amp\`ere's laws
\begin{equation}
\label{AmpereFaraday}
\oint_C {\bf E} \cdot {\bf dl} = - \frac{1}{c} \frac{d}{dt}\int_{S} {\bf B} \cdot {\bf ds} \;\;\;\;\;\;, \;\;\;\;\;\;  \oint_C {\bf H} \cdot {\bf dl} = \frac{1}{c} \int_{S} {\bf J} \cdot {\bf ds} \, ,
\end{equation}
we see that the important quantities in these equations are the {\it flux} of ${\bf B}$ and the {\it circulation} of ${\bf H}$. This clearly suggests that ${\bf B}$ is to be averaged over a surface while ${\bf H}$ should be averaged over a line. For a periodic system we need to average over an unit cell only, so, using the unit cell defined in Fig. 3 and the fact that $H_T$ points in the z-direction, we conclude that a sensible definition of ${\bf B}_{ave}$ and ${\bf H}_{ave}$ is \cite{Pendry, PendrySmith}
\begin{eqnarray}
\label{Bave}
{\bf B}_{\rm ave} = \frac{1}{b^2}\int_{S_z} {\bf B} \cdot {\bf ds}=\frac{1}{b^2}\int_{S_z} {\bf H}_T \cdot {\bf ds} = H   \\
\label{Have}
{\bf H}_{\rm ave}  = \frac{1}{b} \int_{L_z} {\bf H}_T \cdot {\bf dl} = H_T^{\rm out} \, ,
\end{eqnarray}
where $S_z$ is either of the two faces perpendicular to the $z$-axis, and $L_z$ is one of the edges parallel to $\hat{z}$. Now, following \cite{Pendry} and defining the effective magnetic permeability as
\begin{equation}
\label{MuEff}
\mu_{\rm eff}(\omega) = \frac{|\bf B_{ave}|}{|{\bf H_{ave}|}}
\end{equation} 
we get finally
\begin{equation}
\label{MuEff2}
\mu_{\rm eff}(\omega) = \frac{H}{H_T^{\rm out}} = 1 - \frac{\pi r^2}{b^2} \frac{\omega^2}{\omega^2 - c^2/\pi r^2 {\cal C} + 2i\alpha\omega c^2/r} = 1 - f \frac{\omega^2}{\omega^2 - \omega_m^2+ 2i\gamma_m\omega} \, ,
\end{equation} 
where we defined of the filling factor $f$, the resonant frequency $\omega_m$ and the dissipation coefficient $\gamma_m$. 

\begin{figure}[h]
\scalebox{0.25}{\includegraphics{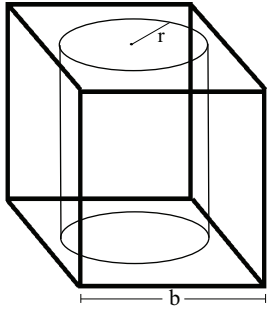}}\hspace{2pc}
\begin{minipage}{14pc}
\vspace{-50pt}
\caption{\small Illustration of the unit cell used in our array. Again, for the sake of simplicity, the cylinder is shown as a single sheet.}
\end{minipage}
\end{figure}

There are some peculiarities about the previous formula that we should point out. The first thing is the functional resemblance between expression (\ref{MuEff2}) and the Drude-Lorentz model for the electric permittivity, the only difference between them being the presence of a $\omega^2$ in the numerator of (\ref{MuEff2}). As we shall see in the next section, this leads to interesting consequences in the Casimir force. In addition, it is easy to see that $\mu_{\rm eff}$ does not go to unity in the high frequency limit, as we should expect on physical grounds. This tells us that (\ref{MuEff2}) cannot be valid up to arbitrarily high frequencies, what makes perfect sense given that all our calculations are based on an effective medium picture that breaks down for very small wavelengths. Our final remark is that, despite the problems in the high frequency range, expression (\ref{MuEff2}) is totally acceptable from causality requirements, since it is analytic in the upper half plane of complex frequencies. Indeed, it may be shown that its real and imaginary part are linked together by a slightly modified Kramers-Kronig formulae  
\begin{eqnarray}
\label{KK1}
&& \mu'(\omega) = (1-f) + \frac{2}{\pi} {\rm P} \!\! \int_0^{\infty} \!\!\! dy y \frac{\mu''(y) - 1}{y^2 - \omega^2} \, ,\nonumber \\
&&\mu''(\omega) =  -\frac{2\omega}{\pi} {\rm P} \!\! \int_0^{\infty} \!\!\! dy \frac{\mu'(y) - (1-f)}{y^2 - \omega^2}
\end{eqnarray}
where $\mu'(\omega)={\rm Re}\,\mu(\omega)$ and $\mu''(\omega)={\rm Im}\,\mu(\omega)$.

\section{The Casimir Force}

Now that we have an explicit formula for the magnetic activity of our MM, we may proceed with the evaluation of the Casimir force. There is, however, an important issue we have to deal with before going directly to calculations. The metamaterial described in the previous section is severely anisotropic, and (\ref{MuEff2}) gives the right magnetic response only when {$\bf H$} is parallel to the $z$-axis. As the Casimir force depends on virtual fluctuations coming from all directions, the inevitable conclusion is that $\mu_{\rm eff}$ has to be generalized in order to take arbitrary propagation directions into account. Unfortunately, the simplicity present in the previous section disappears when we depart from the particular case where ${\bf H} //  \hat{z}$, with effects like polarization mixing and spatial dispersion \cite{Silveirinha} coming into play. In addition, this high degree of complication obscures all the physics we are trying to analyze, namely, magnetic effects affecting the Casimir force. So, in order to avoid unnecessary distractions, from now on we consider a toy model that is based on the one we have been discussing, but with an isotropic magnetic permeability described by eq. (\ref{MuEff2}) and an isotropic electric permittivity given by
\begin{eqnarray}
\label{Drude-Lorentz-Epsilon}
\epsilon_{\rm eff}(\omega) = 1- \frac{\Omega_{e}^2}{\omega^2 - \omega_{e}^2 + i \gamma_{e} \omega},
\end{eqnarray}
where $\Omega_{e}$, $\omega_{e}$ and $\gamma_{e}$ are respectively the electric oscillating strength, resonance frequency and dissipation coefficient. 

The Casimir pressure between two or more bodies can be calculated in fairly general situations using techniques such as the scattering formalism \cite{PauloEmig}, the Krein formula \cite{Wirzba}, the argument principle \cite{DiegoMarachevsky}, path integrals \cite{KlichBordag}, and others. A thorough discussion of these formalisms is outside the scope of this paper, but for our purposes it suffices to say that the result obtained for the pressure is that it depends basically on the reflection properties of the bodies and the distances among them. For the particular case of two real (i.e., non-ideal) isotropic half-spaces separated by vacuum, as shown in Fig. 4, the pressure is given by 
\begin{equation}
\label{Lifshitz}
\frac{F(d)}{A}= 2 \hbar \int_0^{\infty} \hspace{-2pt}  \frac{d\xi}{2 \pi} 
\int  \frac{d^2 {\bf k}_{\|}}{(2 \pi)^2}
\, K_3 \hspace{-4pt} \sum_{j={\rm TE,TM}} \frac{R_1^j \, R_2^j \, e^{-2K_3 d} }{
1- R_1^j \, R_2^j \, e^{-2K_3 d}} ,
\end{equation}
where $K_3=\sqrt{k^2_{\|}+ \xi^2/c^2}$ and the reflection coefficients $R_n^{\rm TE}$, $R_n^{\rm TM}$, are given by
\begin{eqnarray}
\label{FresnelCoefficientsMM}
&&R^{{\rm TE}}_{n}(i\xi, {\bf k}_{\|}) = \frac{\mu_n (i \xi) K_3 - \sqrt{k_{\|}^2 + \mu_n (i \xi)\epsilon_n(i \xi) \xi^2/c^2}}{\mu_n (i \xi)K_3 + \sqrt{k_{\|}^2 + \mu_n (i \xi)\epsilon_n(i \xi) \xi^2/c^2}} \nonumber \\ 
&&R^{{\rm TM}}_{n}(i\xi, {\bf k}_{\|}) = \frac{\epsilon_n (i \xi) K_3 - \sqrt{k_{\|}^2 + \mu_n (i \xi)\epsilon_n(i \xi) \xi^2/c^2}}{\epsilon_n(i \xi) K_3 + \sqrt{k_{\|}^2 +  \mu_n (i \xi)\epsilon_n (i \xi) \xi^2/c^2}} \, ,
\end{eqnarray}
where $\epsilon_n$ and $\mu_n$ are respectively the permittivity and permeability of the n-th half-space. Expression (\ref{Lifshitz}) is slightly more general than the original Lifshitz formula \cite{Lifshitz}, to which it reduces in the case of $\mu_n = 1$.  

In our conventions, a positive (negative) value of the force means attraction (repulsion). We are interested in the case where one of the half-spaces (see fig. 4) is purely metallic, implying in
\begin{eqnarray}
\label{Drude}
\epsilon_1(\omega) = 1 - \frac{\Omega_1^2}{(\omega^2 + i \gamma_1 \omega)} \;\;\;\;\; , \;\;\;\;\;  \mu_1(\omega) = 1 \, ,
\end{eqnarray}
where $\Omega_1$ is the metallic plasma frequency and $\gamma_1$ is the dissipation coefficient, and the other half-space is a MM characterized by $\epsilon_2(\omega) = \epsilon_{\rm eff}(\omega)$ and $\mu_2(\omega) = \mu_{\rm eff}(\omega)$, given respectively by (\ref{Drude-Lorentz-Epsilon}) and (\ref{MuEff}). 

\begin{figure}[h]
\scalebox{0.35}{\includegraphics{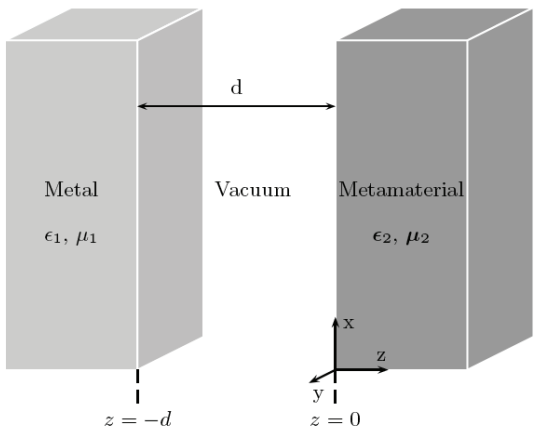}}\hspace{2pc}
\begin{minipage}{14pc}
\vspace{-50pt}
\caption{\small Illustration of the setup used in this paper to calculate the Casimir force.}
\end{minipage}
\end{figure}

Before going straight to the results we would like to point out a small subtlety brought by the use of $\mu_{\rm eff}$ in the Lifshitz formula. Buried in expression (\ref{Lifshitz}) is the assumption that the reflection coefficients are analytic in the whole upper plane of complex frequencies, which is based itself on the assumption that $\epsilon_{1,2}(i\xi) , \mu_{1,2}(i\xi) \geq 0$. It is easy to see that functions $\epsilon_{1,2}(i\xi)$ meet this condition for any choice of parameters, but for $\mu_{2}(i\xi)$ it only holds when $r/b < 1/\sqrt{\pi}$. Fortunately, in our case we have necessarily $r/b < 1/2 < 1/\sqrt{\pi}$ so we don't have to worry, but this little analysis shows that one must be careful with the validity conditions of the Lifshitz formula when considering alternative permittivity and permeability models.
\begin{figure}[h]
\begin{minipage}{14pc}
\scalebox{0.63}{\includegraphics{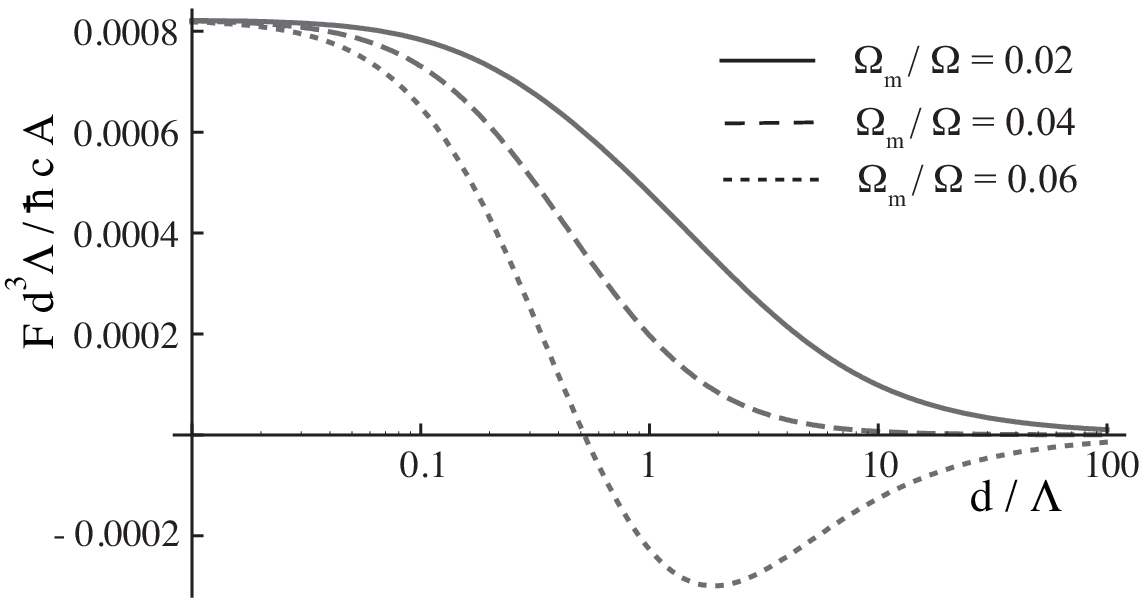}}
\caption{\small The Casimir pressure between a metallic half-space and a MM modeled by permittivity (\ref{Drude-Lorentz-Epsilon}) and permeability (\ref{Drude-Lorentz-Mu}), for several values of the magnetic oscillating strength $\Omega_m$. The parameters used are $\Omega_1/\Omega=0.96$, $\gamma_1/\Omega=0.004$, $\Omega_e/\Omega=0.04$, $\omega_e/\Omega=\omega_m/\Omega=0.1$, 
$\gamma_e/\Omega=\gamma_m/\Omega=0.005$. The frequency scale $\Omega$ is chosen as $1.43 \times 10^{16}$ rad/s and $\Lambda$ is defined as $\Lambda = 2\pi c/\Omega$.}
\end{minipage}\hspace{58pt}
\begin{minipage}{14pc}
\vspace{-44pt}
\scalebox{0.63}{\includegraphics{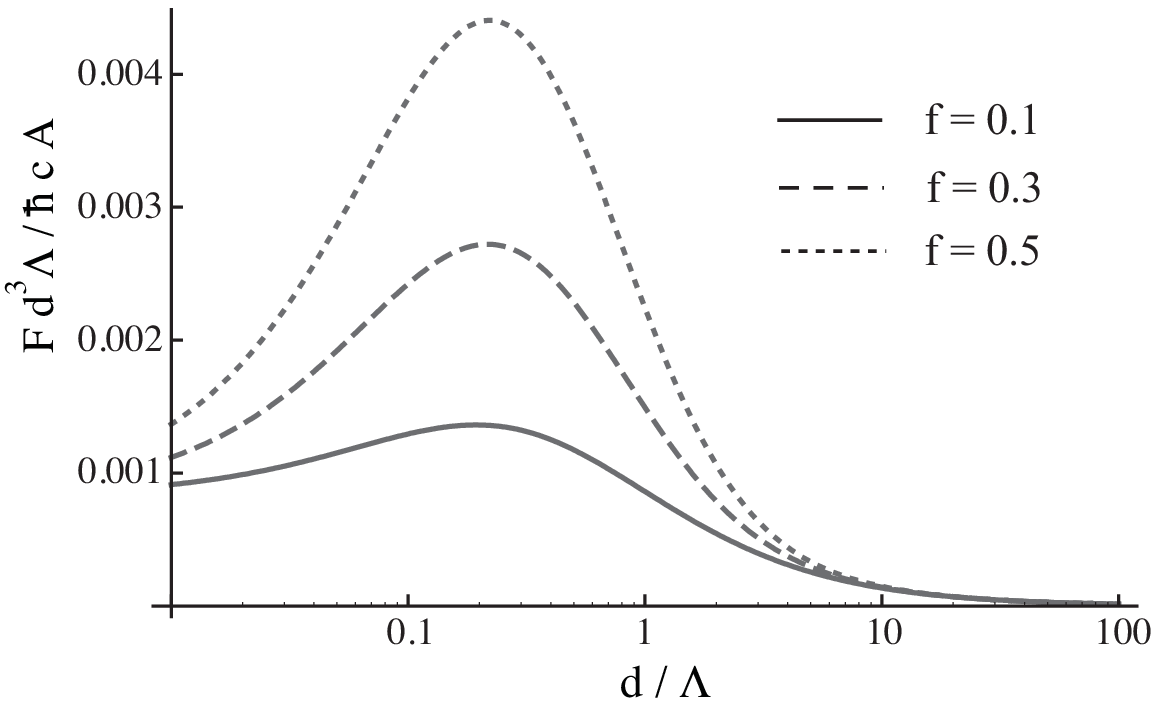}}
\caption{\small The Casimir pressure between a metallic half-space and a MM modeled by permittivity (\ref{Drude-Lorentz-Epsilon}) and permeability (\ref{MuEff2}), for several values of the filling factor $f$. The parameters used are the same as in figure 5.}
\end{minipage} 
\end{figure}

In Figs. 5 and 6 we compare the Casimir pressure as a function of the separation $d$ for different MMs. We see that when the magnetic part is given by a Drude-Lorentz contribution, namely 
\begin{equation}
\label{Drude-Lorentz-Mu} 
\mu_{\rm DL}(\omega) = 1 - \frac{\Omega_{m}^2}{ \omega^2 - 
\omega_{m}^2 + i \gamma_{m} \omega} \, ,
\end{equation}
the force decreases at least as fast as $1/d^3$, while for MMs modeled by (\ref{MuEff2}) we have a transition region where $F/d^3$ actually increases before it goes to zero for large distances. We see also that in the first scenario the force might even assume negative values for strong enough magnetic activity, while on the second case the enhancement of the magnetic part just makes attraction stronger and stronger in the transitional region. This is a clear indication that it is impossible to achieve repulsion with a MM described by (\ref{MuEff2}), and it is actually not difficult to check that this is indeed the case. As we can see from (\ref{Lifshitz}), a necessary (but not sufficient) condition for a negative force to occur is that, for at least one polarization, $R^j_1(i\xi)$ and $R^j_2(i\xi)$ must have different signs. This information is entirely contained in the numerators of both reflection coefficients in (\ref{FresnelCoefficientsMM}), and a simple (but tedious) analysis shows that when the permittivities and permeabilities of the materials involved are given by  (\ref{Drude-Lorentz-Epsilon}), (\ref{Drude}) and (\ref{MuEff2}) we have
\begin{eqnarray}
\label{Unequalities}
&&\mu_n (i \xi) K_3 - \sqrt{k_{\|}^2 + \mu_n (i \xi)\epsilon_n(i \xi) \xi^2/c^2} \; < \; 0 \nonumber \\
&&\epsilon_n (i \xi) K_3 - \sqrt{k_{\|}^2 + \mu_n (i \xi)\epsilon_n(i \xi) \xi^2/c^2} \; >\;  0 \;\;\;\; , \;\;\;\; \forall \; k_{\|} \,,\, \xi \, , \, n \; ,
\end{eqnarray}
what prevents any difference in sign between $R^j_1$ and $R^j_2$. This is to be compared with the case where the MM permeability is described by (\ref{Drude-Lorentz-Mu}), since in this instance it is possible to have $R^{TE}_1(i\xi) \cdot R^{TE}_2(i\xi) < 0$ and therefore Casimir repulsion is not ruled out. We see that a direct consequence of all this reasoning is that repulsion is guaranteed (at least for a given range of distances) provided $\mu(i\xi)$ is significantly larger than $\epsilon(i\xi)$ and that attraction is enforced if $\mu(i\xi) \ll \epsilon(i\xi)$. This usually gives an useful rule of thumb to check if repulsion is possible, or if by changing the relevant permittivities and permeabilities we are approaching (or receding from) Casimir repulsion.  

That two similar models give widely different predictions for the force is a very interesting fact on itself, and, although we just discussed its mathematical roots, a physical explanation for it is still lacking. The problem is that the insertion of magnetic effects in the Casimir effect makes its physical analysis notoriously difficult, even the idealized Boyer setup \cite{Boyer} is not easily grasped from a physical point of view. Part of the explanation for why materials described by either $\mu_{\rm eff}(\omega)$ or $\mu_{\rm DL}(\omega)$ are so differently affected by vacuum fluctuations is probably due to the fact that  the former is a diamagnet for almost all frequencies (except around the resonance), while the latter is a paramagnet up to the vicinities of the resonance (and then becomes a diamagnet), but to go beyond that and state something more conclusive about their distinct physical behavior is, at the current stage, premature.

We would like to end this section with a brief discussion about the effects of electromagnetic anisotropy. As it was discussed to some extent in \cite{usPRA}, anisotropy usually works against repulsion, but in this particular case things may be a bit different. Let us consider an hypothetical MM that is electrically isotropic but has a tensorial magnetic response $\boldsymbol{\mu}$ described by (\ref{MuEff2}) in one direction (say, the $z$ direction) and is magnetically inert in the plane perpendicular to it \footnote{So, in an appropriate base we would have $\mu_{xx}=\mu_{yy}=1$, $\mu_{zz}=\mu_{\rm eff}$}. This kind of MM gives rise to a slightly reduced Casimir attraction if compared to an isotropic MM described by (\ref{MuEff2}), and the reason is that we have $\mu_{xx}(i\xi)/\epsilon_{\rm eff}(i\xi) , \mu_{yy}(i\xi)/\epsilon_{\rm eff}(i\xi) > \mu_{\rm eff}(i\xi)/\epsilon_{\rm eff}(i\xi)$ (since $\mu_{xx}(i\xi)=\mu_{yy}(i\xi)=1>\mu_{\rm eff}(i\xi)$), and hence, by the rule of thumb discussed earlier in this section, we should expect less attraction.

\section{Conclusions}

In this paper we discussed how to obtain the magnetic permeability of a specific MM from a theoretical/phenomenological point of view, and how the Casimir force is affected by it. Unfortunately, the kind of magnetic activity we obtained does not lead to Casimir repulsion, nor even to reduced attraction. Incidentally, these results also show that for Casimir purposes, extrapolations of models for electromagnetic properties must be done in a very careful fashion, as these extrapolations may be crucial to the determination of the Casimir force. Our final conclusion is that, as the description of practical MMs is typically known only on a narrowband of frequencies and incidence angles, much has yet to be learned before a definitive statement about the possibility of repulsive Casimir forces can be made.

\section{Acknowledgements} 

I am greatly indebted to D A R Dalvit and P W Milonni for their encouragement and constructive criticism. I would also like to thank J F O'Hara for useful discussions, and acknowledge the support of the U.S. Department of Energy through the LANL/LDRD program for this work.

\end{document}